\documentclass[12pt]{article}
\usepackage{latexsym,amssymb,amsfonts,amsmath}
\usepackage{mathrsfs}
\usepackage[dvips]{graphicx}
\newlength{\extraspace}
\newlength{\extraspaces}
\newcommand{\be}{\begin{equation}
\addtolength{\abovedisplayskip}{\extraspaces}
\addtolength{\belowdisplayskip}{\extraspaces}
\addtolength{\abovedisplayshortskip}{\extraspace}
\addtolength{\belowdisplayshortskip}{\extraspace}}
\newcommand{\ee}{\end{equation}}

\newcommand{\bea}{\begin{eqnarray}
\addtolength{\abovedisplayskip}{\extraspaces}
\addtolength{\belowdisplayskip}{\extraspaces}
\addtolength{\abovedisplayshortskip}{\extraspace}
\addtolength{\belowdisplayshortskip}{\extraspace}}
\newcommand{\eea}{\end{eqnarray}}

\DeclareMathOperator{\Tr}{Tr}
\DeclareMathOperator{\diag}{diag}

\usepackage{braket}

\newcommand{\I}{\textbf{I}}

\begin{document}

\addtolength{\baselineskip}{.8mm}

{\thispagestyle{empty}

\noindent \hspace{1cm} \hfill IFUP--TH/2025 \hspace{1cm}\\

\begin{center}
\vspace*{1.0cm}
{\Large\bf A new study of the interactions of the axion with mesons and photons using a chiral effective Lagrangian model}\\
\vspace*{1.0cm}
{\large
Enrico Meggiolaro$^{1,~2,~}$\footnote{E-mail: enrico.meggiolaro@unipi.it}
and Mirko Tamburini$^{1}$
}\\
\vspace*{0.5cm}{\normalsize
$^{1}$ {Dipartimento di Fisica, Universit\`a di Pisa,
Largo Pontecorvo 3, I-56127 Pisa, Italy}}\\
\vspace*{0.5cm}{\normalsize
$^{2}$ {INFN, Sezione di Pisa,
Largo Pontecorvo 3, I-56127 Pisa, Italy}}\\
\vspace*{2cm}{\large \bf Abstract}
\end{center}

\noindent
In this paper, we extend the results obtained in a previous work and investigate the most interesting decay processes involving axions, photons and the lightest pseudoscalar mesons in the more general case in which the quarks (and, therefore, the mesons) may be charged under the $U(1)$ Peccei-Quinn symmetry, making use of a chiral effective Lagrangian model with $N_f=3$ light quark flavors, which also includes the flavor-singlet pseudoscalar meson and implements the $U(1)$ axial anomaly of the fundamental theory. In particular, we compute the axion mass, the electromagnetic coupling of the axion to photons, and the amplitudes and widths of the decay processes $\eta/\eta'\rightarrow \pi\pi a$.
}
\newpage

\section{Introduction}
\setcounter{equation}{0}

The so-called ``strong-CP problem'' (i.e., the absence of CP violation in the strong interactions) is one of the open issues of the Standard Model (see, e.g., Refs. \cite{Weinberg-book,VP2009}). Among the several possible solutions, the most appealing is surely the one proposed by Peccei and Quinn (PQ) in 1977 \cite{PQ1977a,PQ1977b} and developed by Weinberg and Wilczek in 1978 \cite{Weinberg1978,Wilczek1978}.
The key idea (see also Ref. \cite{Peccei2008} for a recent review) is to extend the Standard Model by adding a new pseudoscalar particle, called ``axion'', in such a way that there is a new $U(1)$ global symmetry, referred to as $U(1)_{PQ}$, which is both spontaneously broken at a scale $f_a$ and anomalous (i.e., broken by quantum effects), with the related current satisfying the relation $\partial_{\mu}J^{\mu}_{PQ}=a_{PQ}Q$ , where $Q=\frac{g^{2}}{64\pi^{2}}\varepsilon^{\mu\nu\rho\sigma} G^{a}_{\mu\nu}G^{a}_{\rho\sigma}$ is the so-called \emph{topological charge density} and $a_{PQ}$ is the so-called \textit{color anomaly} coefficient.

In the original Peccei-Quinn-Weinberg-Wilczek (PQWW) model \cite{PQ1977a,PQ1977b,Weinberg1978,Wilczek1978} the scale $f_a$ was identified with the electroweak breaking scale $v \approx 250$ GeV, but this leads to large couplings between the axion and the Standard Model fields, which have been ruled out by experiments (see, for example, Ref. \cite{MS2015}).
In order to bypass these experimental bounds, the so-called ``invisible axion'' models were developed, such as the Kim-Shifman-Vainshtein-Zakharov (KSVZ) model \cite{Kim79,SVZ} and the Dine-Fischler-Srednicki-Zhitnisky (DFSZ) model \cite{DFS,Zhitnisky1980}, in which new heavy quarks or scalar fields, charged under $U(1)_{PQ}$ but neutral with respect to the Standard Model gauge group, are introduced. In these models, the $U(1)_{PQ}$ breaking scale $f_{a}$ is a free parameter of the theory and, assuming $f_{a} \gg v$, a very light axion with small couplings to the Standard Model fields is predicted, a scenario which is still compatible with the experimental bounds. At present, the more precise bounds on the $U(1)_{PQ}$ breaking scale come from astrophysical and cosmological considerations (see, for example, Refs. \cite{bounds_a1,bounds_a2,bounds_b}): $10^{9}~ \text{GeV} \lesssim f_a \lesssim 10^{17}~ \text{GeV}$.

In this paper, we shall extend the results obtained in Ref. \cite{LM2020} and investigate the most interesting decay processes involving axions, photons and the lightest pseudoscalar mesons in the more general case in which the quarks (and, therefore, the mesons) may be charged under the $U(1)$ Peccei-Quinn symmetry, making use of a chiral effective Lagrangian model with $N_f=3$ light quark flavors, which also includes the flavor-singlet pseudoscalar meson and implements the $U(1)$ axial anomaly of the fundamental theory.

The chiral effective Lagrangian model considered in Ref. \cite{LM2020} was a simple ``axionized'' version (studied also in Refs. \cite{DS2014,DRVY2017}) of the chiral effective Lagrangian model proposed by Witten, Di Vecchia, Veneziano, \textit{et al.} \cite{Witten80,DV1980,etal80a,etal80b,etal80c,etal80d}, which describes the Nambu-Goldstone bosons originated by the spontaneous breaking of the $SU(3)_L\otimes SU(3)_R$ chiral symmetry (with $N_f=3$ light quark flavors) and the flavor-singlet pseudoscalar meson, implementing the $U(1)$ axial anomaly of the fundamental theory.
This model reproduces in the low-energy effective theory only a class of high-energy axion models (corresponding essentially to the KSVZ model and its variants), where the Standard Model quarks are all \emph{neutral} under the $U(1)_{PQ}$ symmetry transformation and the $U(1)_{PQ}$ anomaly comes from heavy (``exotic'') quarks which have nonzero Peccei-Quinn charges (and which are integrated out in the low-energy effective theory).
However, there exists another class of axion models (like the above-mentioned DFSZ model and its variants, as well as the PQWW model and its variants \cite{Peccei2008,BPY1987}), where also the Standard Model quarks are \emph{charged} under the $U(1)_{PQ}$ symmetry and each of them transforms with a $U(1)$ phase that depends on its PQ charge.

At the beginning of Sect. 2, for the benefit of the reader, we shall briefly recall the chiral effective Lagrangian model considered in Refs. \cite{LM2020,DS2014,DRVY2017}.
Then, we shall propose a generalization of this model in which the light quarks (and, therefore, the mesons) may be charged under a $U(1)_{PQ}$ transformation (each of them transforming with a $U(1)$ phase that depends on its PQ charge).

Using this generalized model, we shall compute the axion mass (in Sect. 3) and the axion-photon-photon coupling constant $g_{a\gamma\gamma}$ (in Sect. 4),
and (in Sect. 5) we shall study the hadronic decays $\eta/\eta'\rightarrow \pi\pi a$ (which, among all the possibile hadronic decays involving also the axion, are the ones involving the lowest-energy hadrons): the results obtained for the various quantities will be compared with the corresponding results found in Ref. \cite{LM2020} and in the literature.

Finally, in Sect. 6 we shall summarize and critically comment on the results that we have obtained in the previous sections for the axion mass and for the electromagnetic and the hadronic processes involving the axion, focusing in particular on the question of their dependence (or independence) on the details of the axion model considered.

\section{The effective Lagrangian model of Witten, Di Vecchia, Veneziano, \emph{et al.}, with the inclusion of the axion: a generalization}
\setcounter{equation}{0}

The effective Lagrangian model proposed by Witten, Di Vecchia, Veneziano, \textit{et al.} \cite{Witten80,DV1980,etal80a,etal80b,etal80c,etal80d} describes the Nambu-Goldstone bosons originated by the spontaneous breaking of the $SU(3)_L\otimes SU(3)_R$ chiral symmetry and the flavor-singlet pseudoscalar meson, implementing the $U(1)$ axial anomaly of the fundamental theory. We will refer to it as the ``WDV model''. Even if this model was derived and fully justified in the large-$N_c$ limit ($N_c$ being the number of colors), the numerical results obtained for the physical value $N_c=3$ turn out to be quite consistent with experimental data.
In Refs. \cite{LM2020,DS2014,DRVY2017} the following ``axionized'' version of the WDV model was considered, including the axion field $S_a$ in addition to the meson fields:
\begin{equation}\label{WDV+axion_old}
\begin{split}
\mathcal{L} = &\frac{F_\pi^2}{4}\Tr\left[\partial_{\mu}U\partial^{\mu}U^{\dagger}\right] + \frac{f_a^2}{2}\partial_{\mu}N\partial^{\mu}N^{\dagger} + \frac{BF_\pi^2}{2}\Tr\left[M(U+U^{\dagger})\right] \\
& + \frac{i}{2}Q\left\{\Tr[\ln U-\ln U^{\dagger}] + a_{PQ}(\ln N-\ln N^{\dagger})\right\} + \frac{Q^{2}}{2A} ,
\end{split}
\end{equation}
where $N = e^{i\frac{S_a}{f_a}}$ parametrizes the axion field in the standard notation for Nambu-Goldstone bosons, $f_a$ being the energy scale at which the new $U(1)_{PQ}$ global symmetry is broken.\\
Concerning the meson fields, the usual nonlinear parametrization is adopted in terms of the following $3\times 3$ unitary complex matrix:
\begin{equation}\label{U-field}
U(x) = e^{\frac{i}{F_\pi}\Phi(x)} ,~~\text{with}~~
\Phi(x) = \sum_{a=1}^{8}\pi_{a}(x)\lambda_{a}+\sqrt{\frac{2}{3}}S(x)\I ,
\end{equation}
where $\lambda_a ~(a=1,\ldots,8)$ are the usual generators of $SU(3)$ (\textit{Gell-Mann matrices}), normalized so as $\Tr\left[\lambda_a\lambda_b\right]=2\delta_{ab}$, and $\pi_{a}(x) $ are the nonsinglet pseudoscalar-meson fields, while $S(x)$ is the flavor-singlet pseudoscalar-meson field.
Moreover:
\begin{itemize}
\item $F_\pi$ is the pion decay constant.
\item $M = \diag(m_u,m_d,m_s)$ is the quark-mass matrix.
\item B is a constant (with the dimension of a mass) which relates the squared masses of the pseudoscalar mesons and the quark masses: for example, $m_{\pi}^{2}=B(m_{u}+m_{d})$.
\end{itemize}
The topological charge density $Q(x)$
is introduced as an auxiliary field, whereas $A$ is a parameter which (at least in the large-$N_c$ limit) can be identified with the topological susceptibility in the pure Yang-Mills theory ($A = -i \int d^4 x \langle T Q(x) Q(0) \rangle\vert_{YM}$).

Under a general $SU(3)_L\otimes SU(3)_R\otimes U(1)_A$ transformation, the quark fields in the QCD Lagrangian transform as $q_L \to q'_L = \widetilde{V}_L q_L$ and $q_R \to q'_R = \widetilde{V}_R q_R$, where $\widetilde{V}_{L}=e^{i\beta}V_L$, $\widetilde{V}_{R}=e^{-i\beta}V_R$, with ${V}_{L,R}\in SU(3)$,\footnote{Throughout this paper, we shall use the following notations for the left-handed and right-handed quark fields: $q_{L,R} \equiv \frac{1}{2}(1 \pm \gamma_{5})q$, with $\gamma_5 \equiv -i\gamma^0\gamma^1\gamma^2\gamma^3$. Moreover, we shall adopt the convention $\varepsilon^{0123} = -\varepsilon_{0123} = 1$ for the (Minkowskian) completely antisymmetric tensor $\varepsilon^{\mu\nu\rho\sigma}$ ($=-\varepsilon_{\mu\nu\rho\sigma}$) which appears in the expressions of the topological charge density $Q$ and of the dual electromagnetic field-strength tensor $\tilde{F}^{\mu\nu}$.}
while the meson field matrix $U_{ij}$ in the effective Lagrangian transforms as the quark bilinear matrix $\bar{q}_{jR}q_{iL}$, that is
\begin{equation}\label{U-transformation}
U \rightarrow U' = \widetilde{V}_{L}U\widetilde{V}_{R}^{\dagger} .
\end{equation}
It is easy to see that the $U$-dependent anomalous term in Eq. \eqref{WDV+axion_old}, that is
\begin{equation}\label{U-anomaly}
\Delta\mathcal{L}_{U,\text{anomaly}} = \frac{i}{2} Q \Tr[\ln U - \ln U^{\dagger}] = -\sqrt{6} Q \frac{S}{F_\pi} ,
\end{equation}
is invariant under $SU(3)_L\otimes SU(3)_R$, while under $U(1)_A$ ($U \to U' = e^{2i\beta}U$, that is $S \to S' = S + \beta \sqrt{6} F_\pi$) it transforms as
\begin{equation}
U(1)_A:~~~~ \Delta\mathcal{L}_{U,\text{anomaly}}\rightarrow \Delta\mathcal{L}_{U,\text{anomaly}} - 6\beta Q ,
\end{equation}
so correctly reproducing the $U(1)$ axial anomaly of the fundamental theory ($\partial_{\mu}J^{\mu}_{5} = 6Q$ in the chiral limit $M=0$, $J^{\mu}_{5}$ being the $U(1)$ axial current).\\
All the terms in the Lagrangian \eqref{WDV+axion_old} are invariant under a $U(1)_{PQ}$ transformation,
\begin{equation}\label{u1pq_old}
U(1)_{PQ}:~~~~ S_a \rightarrow S_a'= S_a + \gamma f_{a} ~~~~ (\text{that is}:~ N \rightarrow N' = N e^{i\gamma}) ,
\end{equation}
apart from the $N$-dependent anomalous term, that is
\begin{equation}\label{N-anomaly}
\Delta\mathcal{L}_{N,\text{anomaly}} = \frac{i}{2} a_{PQ} Q (\ln N - \ln N^{\dagger}) = -a_{PQ} Q \frac{S_a}{f_a} ,
\end{equation}
which instead transforms as
\begin{equation}
U(1)_{PQ}:~~~~ \Delta\mathcal{L}_{N,\text{anomaly}}\rightarrow \Delta\mathcal{L}_{N,\text{anomaly}} - \gamma a_{PQ} Q ,
\end{equation}
so correctly reproducing the $U(1)_{PQ}$ \emph{color} anomaly ($\partial_{\mu}J^{\mu}_{PQ} = a_{PQ} Q$, $J^{\mu}_{PQ}$ being the $U(1)_{PQ}$ current and $a_{PQ}$ being the so-called \emph{color anomaly} coefficient).

This ``axionized'' version of the WDV model was employed in Ref. \cite{LM2020} to study the axion mass and its low-energy interactions with photons and mesons.\\
This model reproduces in the low-energy effective theory only a class of high-energy axion models (corresponding essentially to the KSVZ model and its variants), where the Standard Model quarks are all \emph{neutral} under the $U(1)_{PQ}$ symmetry transformation and the $U(1)_{PQ}$ anomaly comes from heavy (``exotic'') quarks which have nonzero Peccei-Quinn charges (and which are integrated out in the low-energy effective theory).
However, there exists another class of axion models (like the above-mentioned DFSZ model and its variants, as well as the PQWW model and its variants), where also the Standard Model quarks are \emph{charged} under the $U(1)_{PQ}$ symmetry and each of them transforms with a $U(1)$ phase that depends on its PQ charge.\\
We then propose a generalization of this model that implements a more general version of the $U(1)_{PQ}$ symmetry, in which the $U(1)_{PQ}$ transformation acts not only on the axion field, but also on the three light quarks, precisely as a chiral transformation of the \emph{right-handed} quarks $q_{f,R}$, each one with a different $U(1)$ phase depending on its Peccei-Quinn charge $p_f$. Therefore, in place of Eq. \eqref{u1pq_old}, we now have:
\begin{equation}\label{u1pq_new}
U(1)_{PQ}:~~~~ S_a \rightarrow S_a'= S_a + \gamma f_a , ~~~~
q_R \rightarrow q'_R = e^{-i\gamma\mathcal{Q}_{PQ}} q_R ,
\end{equation}
where $\mathcal{Q}_{PQ}$ is a real and diagonal matrix containing the PQ charges of the three light quarks: $\mathcal{Q}_{PQ} = \diag(p_u, p_d, p_s)$; that is, in terms of the fields appearing in the effective Lagrangian [recalling that the meson field matrix $U_{ij}$ transforms as the quark bilinear matrix $\bar{q}_{jR}q_{iL}$, see Eq. \eqref{U-transformation}]:
\begin{equation}\label{U(1)_PQ_new}
U(1)_{PQ}:~~~~ N \rightarrow N' = N e^{i \gamma} , ~~~~
U \rightarrow U' = U e^{i\gamma\mathcal{Q}_{PQ}} .
\end{equation}
Eq. \eqref{u1pq_new} just mimics the implementation of the $U(1)_{PQ}$ symmetry considered in the PQWW and DFSZ axion models and their variants.\\
In principle, one may also consider a different definition of the $U(1)_{PQ}$ symmetry considering, instead, a chiral transformation of the \emph{left-handed} quark fields,
$q_L \rightarrow q'_L = e^{i\gamma\mathcal{Q}_{PQ}} q_L$ (i.e., with $\tilde{V}_L = e^{i\gamma\mathcal{Q}_{PQ}}$ and $\tilde{V}_R = 0$),
and, therefore, $U \rightarrow U' = e^{i\gamma\mathcal{Q}_{PQ}} U$,
or also an \emph{axial} transformartion of the quark fields,
$q \rightarrow q' = e^{i\gamma\frac{\mathcal{Q}_{PQ}}{2}\gamma_5} q$, that is to say
$q_L \rightarrow q'_L = e^{i\gamma\frac{\mathcal{Q}_{PQ}}{2}} q_L$ and
$q_R \rightarrow q'_R = e^{-i\gamma\frac{\mathcal{Q}_{PQ}}{2}} q_R$
(i.e., with $\tilde{V}_L = \tilde{V}_R^\dagger = e^{i\gamma\frac{\mathcal{Q}_{PQ}}{2}}$),
and, therefore, $U \rightarrow U' = e^{i\gamma\frac{\mathcal{Q}_{PQ}}{2}} U e^{i\gamma\frac{\mathcal{Q}_{PQ}}{2}}$:
they are just equivalent ways to implement the aforementioned symmetry, as we shall explicitly verify below.\\
Therefore, we need to modify the ``axionized'' WDV Lagrangian in Eq. \eqref{WDV+axion_old} in order to implement the aforementioned $U(1)_{PQ}$ symmetry.\\
The first two (kinetic) terms in Eq. \eqref{WDV+axion_old} are, of course, still invariant under the global symmetry transformation \eqref{U(1)_PQ_new}, while the mass term is not and must be modified as follows, in order to make it invariant:
\begin{equation}\label{mass-term}
\Delta\mathcal{L}_{M} = \frac{B F_{\pi}^2}{2} \Tr \big[M (\tilde{U} + \tilde{U}^{\dagger})\big] , 
\end{equation}
having defined:
\begin{equation}\label{U-tilde}
\tilde{U} \equiv U e^{-i\mathcal{Q}_{PQ} \frac{S_a}{f_a}} ,
\end{equation}
which is invariant under the transformation \eqref{U(1)_PQ_new}.\\
Let us observe that, being $M$ and $\mathcal{Q}_{PQ}$ diagonal (and so commuting) matrices and making use of the cyclic property of the trace, one can also rewrite the mass term \eqref{mass-term} replacing $\tilde{U}$ with
$\hat{U} \equiv e^{-i\mathcal{Q}_{PQ} \frac{S_a}{f_a}} U$,
which is invariant under the \emph{left-handed} version of the $U(1)_{PQ}$ transformation, i.e., $S_a \rightarrow S_a'= S_a + \gamma f_a$ and
$U \rightarrow U' = e^{i\gamma\mathcal{Q}_{PQ}} U$,
or also with
$\overline{U} \equiv e^{-i \frac{\mathcal{Q}_{PQ}}{2} \frac{S_a}{f_a}} U e^{-i \frac{\mathcal{Q}_{PQ}}{2} \frac{S_a}{f_a}}$,
which is invariant under the \emph{axial} version of the $U(1)_{PQ}$ transformation, i.e., $S_a \rightarrow S_a'= S_a + \gamma f_a$ and
$U \rightarrow U' = e^{i\gamma\frac{\mathcal{Q}_{PQ}}{2}} U e^{i\gamma\frac{\mathcal{Q}_{PQ}}{2}}$.\\
Alternatively, we can also recast this term in the following way:
\begin{equation}\label{mass-term_bis}
\Delta\mathcal{L}_{M} = \frac{B F_{\pi}^2}{2} \Tr \big[\tilde{M} U + \tilde{M}^\dagger U^\dagger \big] ,
\end{equation}
having defined:
$\tilde{M} \equiv \diag \left( m_u e^{-i p_u S_a/f_a}, m_d e^{-i p_d S_a/f_a}, m_s e^{-i p_s S_a/f_a} \right)$.\footnote{We notice that the form of the modified mass term just mimics the Yukawa Lagrangian in the PQWW and DFSZ models, where the axion is introduced as a phase between the Higgs fields.}\\
Also the anomalous term proportional to $Q$ in Eq. \eqref{WDV+axion_old} must be modified in order to correctly reproduce both the $U(1)$ axial anomaly and the $U(1)_{PQ}$ color anomaly of the fundamental theory.
In fact, let us observe that now the $U$-dependent anomalous term \eqref{U-anomaly} is no more invariant under the $U(1)_{PQ}$ transformation \eqref{U(1)_PQ_new}, but it changes as
\begin{equation}
U(1)_{PQ}:~~~~ \Delta\mathcal{L}_{U,\text{anomaly}}\rightarrow \Delta\mathcal{L}_{U,\text{anomaly}} - \gamma \Tr\mathcal{Q}_{PQ} \, Q .
\end{equation}
As a consequence, observing also that the following relation holds:
\begin{equation}\label{relation-for-utilde}
\Tr[\ln \tilde{U} - \ln \tilde{U}^{\dagger}] = \Tr[\ln U - \ln U^{\dagger}] - \Tr\mathcal{Q}_{PQ} \, (\ln N - \ln N^{\dagger}) ,
\end{equation}
the anomalous term must be modified as follows:
\begin{equation}\label{anomalous-term}
\begin{split}
\Delta\mathcal{L}_{\text{anomaly}} & = \frac{i}{2} Q \big\{ \Tr[\ln \tilde{U} - \ln \tilde{U}^{\dagger}] + a_{PQ} (\ln N - \ln N^{\dagger}) \big\} \\
& = \frac{i}{2} Q \big\{ \Tr[\ln U - \ln U^{\dagger}] + \tilde{a}_{PQ} (\ln N - \ln N^{\dagger}) \big\} ,
\end{split}
\end{equation}
where $\tilde{a}_{PQ}$ is a new color anomaly coefficient defined as:
\begin{equation}\label{a-tilde_PQ}
\tilde{a}_{PQ} \equiv a_{PQ} - \Tr\mathcal{Q}_{PQ} = a_{PQ} - (p_u + p_d + p_s) .
\end{equation}
This quantity (that we could call the \textit{reduced} color anomaly coefficient) represents the contribution to the $U(1)_{PQ}$ anomaly coming only from the \emph{heavy} (both Standard Model and ``exotic'') degrees of freedom, which are integrated out in this low-energy effective model. By contrast, $a_{PQ}$ is the \emph{total} color anomaly coefficient, that takes into account the $U(1)_{PQ}$ anomaly coming from \emph{all} (both Standard Model and ``exotic'') degrees of freedom, including the three light quarks.\\
Analogously to what we have already observed for the mass term \eqref{mass-term}, we note that (making use of the relation $\Tr [\ln H] = \ln (\det H)$ and of the well-known properties of the determinant) also the relation \eqref{relation-for-utilde} and so also the anomalous term \eqref{anomalous-term} can be rewritten replacing $\tilde{U}$ with
$\hat{U} \equiv e^{-i\mathcal{Q}_{PQ} \frac{S_a}{f_a}} U$,
which is invariant under the \emph{left-handed} version of the $U(1)_{PQ}$ transformation, i.e., $S_a \rightarrow S_a'= S_a + \gamma f_a$ and
$U \rightarrow U' = e^{i\gamma\mathcal{Q}_{PQ}} U$,
or also with
$\overline{U} \equiv e^{-i \frac{\mathcal{Q}_{PQ}}{2} \frac{S_a}{f_a}} U e^{-i \frac{\mathcal{Q}_{PQ}}{2} \frac{S_a}{f_a}}$,
which is invariant under the \emph{axial} version of the $U(1)_{PQ}$ transformation, i.e., $S_a \rightarrow S_a'= S_a + \gamma f_a$ and
$U \rightarrow U' = e^{i\gamma\frac{\mathcal{Q}_{PQ}}{2}} U e^{i\gamma\frac{\mathcal{Q}_{PQ}}{2}}$.

All things considered, the new ``axionized'' version of the WDV Lagrangian is thus given by:
\begin{equation}\label{WDV+axion_new}
\begin{split}
\mathcal{L} = &\frac{F_\pi^2}{4}\Tr\left[\partial_{\mu}U\partial^{\mu}U^{\dagger}\right] + \frac{f_a^2}{2}\partial_{\mu}N\partial^{\mu}N^{\dagger} + \frac{BF_\pi^2}{2}\Tr\left[M(\tilde{U}+\tilde{U}^{\dagger})\right] \\
& + \frac{i}{2}Q\left\{\Tr[\ln U-\ln U^{\dagger}] + \tilde{a}_{PQ}(\ln N-\ln N^{\dagger})\right\} + \frac{Q^{2}}{2A} .
\end{split}
\end{equation}
The auxiliary field $Q$ can be eventually integrated out by using its equation of motion ($\frac{\delta \mathcal{L}}{\delta Q} = 0$):
\begin{equation}
Q = -\frac{i}{2}A\left\{\Tr[\ln U-\ln U^{\dagger}] + \tilde{a}_{PQ}(\ln N-\ln N^{\dagger})\right\} .
\end{equation}
The resulting Lagrangian is given by
\begin{equation}\label{WDV+axion}
\begin{split}
\mathcal{L} = &\frac{F_\pi^2}{4}\Tr\left[\partial_{\mu}U\partial^{\mu}U^{\dagger}\right] + \frac{f_a^2}{2}\partial_{\mu}N\partial^{\mu}N^{\dagger} + \frac{BF_\pi^2}{2}\Tr\left[M(\tilde{U}+\tilde{U}^{\dagger})\right] \\
& + \frac{A}{8}\left\{\Tr[\ln U-\ln U^{\dagger}] + \tilde{a}_{PQ}(\ln N-\ln N^{\dagger})\right\}^2 .
\end{split}
\end{equation}

\section{Axion mass}
\setcounter{equation}{0}

After expanding the Lagrangian up to the second order in the fields, we find out that the squared-mass matrix for the mesons $\pi_1$, $\pi_2$, $\pi_4$, $\pi_5$, $\pi_6$, $\pi_7$ is already diagonal, with the well-known eigenvalues: $m^2_{\pi_{1,2}} = m^2_{\pi^{\pm}} = B (m_u + m_d)$, $m^2_{\pi_{4,5}} = m^2_{K^{\pm}} = B (m_u + m_s)$, and $m^2_{\pi_{6,7}} = m^2_{K^0, \bar{K}^0} = B (m_d + m_s)$,
while the fields $\pi_3$, $\pi_8$, $S$ and $S_a$ mix together with the following squared-mass matrix:
\begin{equation}\label{squared-mass-matrix}
\mathcal{M}^2 =
\begin{pmatrix}
2B\tilde{m} & \frac{1}{\sqrt{3}}B\Delta & \sqrt{\frac{2}{3}}B\Delta & - \frac{B F_{\pi}}{f_a} m_{p_1} \\
\frac{1}{\sqrt{3}}B\Delta & \frac{2}{3}B \left(\tilde{m} + 2m_s\right) & \frac{2\sqrt{2}}{3}B \left(\tilde{m} - m_s\right) & - \frac{B F_{\pi}}{\sqrt{3} f_a} m_{p_2} \\
\sqrt{\frac{2}{3}}B\Delta & \frac{2\sqrt{2}}{3}B \left(\tilde{m} - m_s\right) & \frac{6A}{F_{\pi}^2} + \frac{2}{3}B\left(2\tilde{m} + m_s \right) & \frac{\sqrt{6} A \tilde{a}_{PQ}}{F_{\pi} f_a} - \sqrt{\frac{2}{3}}\frac{B F_{\pi}}{f_a} m_{p_3} \\
- \frac{B F_{\pi}}{f_a} m_{p_1} & - \frac{B F_{\pi}}{\sqrt{3} f_a} m_{p_2} & \frac{\sqrt{6} A \tilde{a}_{PQ}}{F_{\pi} f_a} - \sqrt{\frac{2}{3}}\frac{B F_{\pi}}{f_a} m_{p_3} & \frac{A \tilde{a}_{PQ}^2}{f_a^2} + \frac{B F_{\pi}^2}{f_a^2} m_{p_4} \\
\end{pmatrix} ,
\end{equation}
where we have defined the following quantities to display the matrix in a more suitable way:
\begin{equation}\label{mtilde-delta}
\tilde{m} \equiv \frac{1}{2} (m_u + m_d) , \ \ \ \Delta \equiv m_u - m_d ,
\end{equation}
and
\begin{equation}\label{Parameters}
\begin{cases}
m_{p_1} \equiv m_u p_u - m_d p_d ,\\
m_{p_2} \equiv m_u p_u + m_d p_d - 2m_s p_s ,\\
m_{p_3} \equiv m_u p_u + m_d p_d + m_s p_s ,\\
m_{p_4} \equiv m_u p_u^2 + m_d p_d^2 + m_s p_s^2 .\\
\end{cases}
\end{equation}
This squared-mass matrix represents a generalization of the one reported in Ref. \cite{LM2020}, which can be easily recovered in the particular case in which the three light quarks are neutral under the $U(1)_{PQ}$ symmetry, i.e., when
$p_u = p_d = p_s = 0$ ($\mathcal{Q}_{PQ} = 0$).\\
The fields $\pi_{3}$, $\pi_{8}$, $S$, $S_a$ can be written in terms of the ``physical'' fields $\pi^0$, $\eta$, $\eta'$, $a$, associated with the mass eigenstates of Eq. \eqref{squared-mass-matrix}, as follows:
\begin{equation}\label{mixing}
\bigg .\left(
\begin{array}{c}
\pi_{3} \\
\pi_{8} \\
S \\
S_a \\
\end{array}
\bigg .\right)
=
\bigg .\left(
\begin{array}{cccc}
\theta_{\pi_{3}\pi_{3}} & \theta_{\pi_{3}\pi_{8}} & \theta_{\pi_{3}S} & \theta_{\pi_{3}S_a} \\
\theta_{\pi_{8}\pi_{3}} & \theta_{\pi_{8}\pi_{8}} & \theta_{\pi_{8}S} & \theta_{\pi_{8}S_a} \\
\theta_{S\pi_{3}} & \theta_{S\pi_{8}} & \theta_{SS} & \theta_{SS_a} \\
\theta_{S_a\pi_{3}} & \theta_{S_a\pi_{8}} & \theta_{S_aS} & \theta_{S_aS_a}
\end{array}
\bigg .\right)
\bigg .\left(
\begin{array}{c}
\pi^{0} \\
\eta \\
\eta' \\
a
\end{array}
\bigg .\right) ,
\end{equation}
where $\theta_{ij}$ is an orthogonal mixing matrix.

Our goal is now to compute the axion mass, which corresponds to the lightest eigenvalue of the matrix \eqref{squared-mass-matrix}.
From the astrophysical and cosmological bounds on the scale $f_a$ \cite{bounds_a1,bounds_a2,bounds_b} (or better on $f_a/a_{PQ}$, but $a_{PQ} \sim \mathcal{O}(1)$ for the more realistic axion models \cite{DMN2017}) we have: $10^{9}~\textrm{GeV} \lesssim f_a \lesssim 10^{17}~\textrm{GeV}$. As a consequence, it is surely legitimate to perform the computation only at the first nontrivial order in an expansion in powers of $1/f_a$ (that is to say, of $F_\pi/f_a$).
Exploiting this fact and considering that the determinant of $\mathcal{M}^2$ is equal to the products of its eigenvalues, we can derive the following expression for the squared mass of the axion at the leading order in $f_a^{-1}$:
\begin{equation}\label{Axion_Mass_Formula}
m^2_a \simeq \frac{\det \mathcal{M}^2}{\det \mathcal{M}^2_<} ,
\end{equation}
where $\mathcal{M}^2_<$ is the upper-left minor obtained from $\mathcal{M}^2$ by removing the last row and the last column:
\begin{equation}
\mathcal{M}^2_< = 
\begin{pmatrix}
2B\tilde{m} & \frac{1}{\sqrt{3}}B\Delta & \sqrt{\frac{2}{3}}B\Delta \\
\frac{1}{\sqrt{3}}B\Delta & \frac{2}{3}B \left(\tilde{m} + 2m_s\right) & \frac{2\sqrt{2}}{3}B \left(\tilde{m} - m_s\right) \\
\sqrt{\frac{2}{3}}B\Delta & \frac{2\sqrt{2}}{3}B \left(\tilde{m} - m_s\right) & \frac{6A}{F_{\pi}^2} + \frac{2}{3}B\left(2\tilde{m} + m_s \right) \\
\end{pmatrix} .
\end{equation}
The expression \eqref{Axion_Mass_Formula} is justified by the fact that the eigenvalues $m^2_{\pi^0}$, $m^2_\eta$, $m^2_{\eta'}$, $m^2_a$ of the squared mass matrix \eqref{squared-mass-matrix} get corrections of order $\mathcal{O}(f_a^{-2})$ with respect to their ``unperturbed'' values $m^2_{\pi^0}|_0$, $m^2_\eta|_0$, $m^2_{\eta'}|_0$, $m^2_a|_0$ in the limit $f_a\to\infty$ (i.e., $m^2_{\pi^0} = m^2_{\pi^0}|_0 + \mathcal{O}(f_a^{-2})$, etc.) and, moreover, the axion turns out to be massless in this limit [i.e., $m^2_a|_0 = 0$, that is to say $m^2_a = \mathcal{O}(f_a^{-2})$] and it is decoupled from the other neutral pseudoscalar mesons in the squared-mass matrix \eqref{squared-mass-matrix}.
Accordingly, the determinant of $\mathcal{M}^2_<$ is equal to the product of the the squared masses of $\pi^0$, $\eta$, and $\eta'$ mesons, in the limit $f_a\to\infty$: $\det \mathcal{M}^2_< = m^2_{\pi^0}|_0 \cdot m^2_{\eta}|_0 \cdot m^2_{\eta'}|_0$.
It can be easily computed, finding the following result:
\begin{equation}\label{det_M2_minor}
\det \mathcal{M}^2_< = \frac{8 A B^2}{F_{\pi}^2} \biggl(m_u m_d + m_u m_s + m_d m_s + \frac{B F_{\pi}^2}{A} m_u m_d m_s\biggr) .
\end{equation}
Rather, in order to evaluate the determinant of $\mathcal{M}^2$, we have used the software \emph{Mathematica}, which has yielded us the outcome:
\begin{equation}\label{det_M2}
\det \mathcal{M}^2 = \frac{8 A B^3}{f_a^2} \Big(\tilde{a}_{PQ} + \text{Tr} \, \mathcal{Q}_{PQ}\Big)^2 m_u m_d m_s = \frac{8 A B^3}{f_a^2} a^2_{PQ} m_u m_d m_s ,
\end{equation}
where in the second passage we have used the definition of $\tilde{a}_{PQ}$ given in Eq. \eqref{a-tilde_PQ}.\\
Then, substituting Eqs. \eqref{det_M2_minor} and \eqref{det_M2}) into Eq. \eqref{Axion_Mass_Formula}, we find out that the squared mass of the axion, at the leading order in $f_a^{-1}$, is given by:
\begin{equation}\label{Axion_Mass}
m^2_a = \left(\frac{a_{PQ}}{f_a}\right)^2 B F_\pi^2 \frac{m_u m_d m_s}{m_u m_d + m_u m_s + m_d m_s + \frac{B F_{\pi}^2}{A} m_u m_d m_s} .
\end{equation}
Recalling the expression for the QCD \emph{topological susceptibility} (defined as\\ $\chi_{QCD} \equiv-i \int d^4 x \langle T Q(x) Q(0) \rangle\vert_{QCD}$)
which is found using the WDV model (see Refs. \cite{DS2014,LM2018} and references therein),
\begin{equation}
\chi_{QCD} = B F_{\pi}^2 \frac{m_u m_d m_s}{m_u m_d + m_u m_s + m_d m_s + \frac{B F_{\pi}^2}{A} m_u m_d m_s} ,
\end{equation} 
Eq. \eqref{Axion_Mass} can be rewritten as
\begin{equation}\label{Axion_Mass_Chi}
m^2_a = \left(\frac{a_{PQ}}{f_a}\right)^2 \chi_{QCD} ,
\end{equation}
thus confirming the well-known relationship (valid at the leading order in $1/f_a$) between the squared mass of the axion and the QCD topological susceptibility \cite{SVZ}.
Therefore, the squared mass of the axion turns out to be model independent, apart from the multiplicative factor $(a_{PQ}/f_a)^2$, which contains the \emph{total} color anomaly coefficient $a_{PQ}$.

\section{Electromagnetic decay of the axion}
\setcounter{equation}{0}

All axion models predict an axion-photon-photon coupling and therefore the electromagnetic decay of the axion in two photons: most of the experimental research concerning the axion is focused on this process (see, for example, Ref. \cite{IR2018} for an exhaustive review on both the theoretical aspects and the experimental research of axions and axion-like particles).

In order to investigate the electromagnetic decay of the axion, we have to introduce the electromagnetic interactions into the Lagrangian \eqref{WDV+axion}. This is done (i) by replacing the derivative of the field $U$ with the corresponding covariant derivative $D_\mu U = \partial_{\mu}U + i e A_\mu[\mathcal{Q}_{e.m.},U]$, where $A_\mu$ is the electromagnetic field and $\mathcal{Q}_{e.m.} = \diag(2/3,-1/3,-1/3)$ is the quark electric-charge matrix (in units of $e$, the absolute value of the electron charge), (ii) by adding the following term, which reproduces the electromagnetic anomaly of the $U(1)$ and $SU(3)$ axial currents (see Ref. \cite{DNPV1981}):
\begin{equation}\label{U-em-anomaly}
\Delta\mathcal{L}_{U,\text{anomaly}}^{\text{(e.m.)}}=\frac{i}{2}G\Tr[\mathcal{Q}_{e.m.}^{2}\left(\ln U -\ln U^{\dagger}\right)] = -\frac{G}{3F_{\pi}}\left(\pi_{3}+\frac{1}{\sqrt{3}}\pi_{8}+\frac{2\sqrt{2}}{\sqrt{3}}S\right) ,
\end{equation}
where $G=\frac{e^2 N_C}{32\pi^2}\varepsilon^{\mu\nu\rho\sigma}F_{\mu\nu}F_{\rho\sigma} = \frac{3 e^2}{16\pi^2} F_{\mu\nu}\tilde{F}^{\mu\nu}$, $F_{\mu\nu}$ being the electromagnetic field-strenght tensor and $\tilde{F}^{\mu\nu}=\frac{1}{2}\varepsilon^{\mu\nu\rho\sigma}F_{\rho\sigma}$ its dual,
and (iii) by also adding the following $N$-dependent anomalous term:
\begin{equation}\label{N-em-anomaly}
\Delta\mathcal{L}_{N,\text{anomaly}}^{\text{(e.m.)}} = \frac{i}{2} \tilde\xi_{PQ} G (\ln N - \ln N^{\dagger}) = -\tilde\xi_{PQ} G \frac{S_a}{f_a} ,
\end{equation}
where
\begin{equation}\label{xi-tilde_PQ}
\tilde\xi_{PQ} \equiv \xi_{PQ} - \Tr[\mathcal{Q}_{e.m.}^2\mathcal{Q}_{PQ}] = \xi_{PQ} - \frac{4 p_u + p_d + p_s}{9} ,
\end{equation}
$\xi_{PQ}$ being the so-called \emph{electromagnetic anomaly coefficient} (so that $\tilde\xi_{PQ}$ could be called the \textit{reduced} electromagnetic anomaly coefficient).
The sum $\Delta\mathcal{L}_{\text{anomaly}}^{\text{(e.m.)}}$ of the two anomalous terms \eqref{U-em-anomaly} and \eqref{N-em-anomaly} thus correctly reproduces also the $U(1)_{PQ}$ electromagnetic anomaly $\xi_{PQ} G$ of the $U(1)_{PQ}$ current: $\partial_{\mu}J^{\mu}_{PQ} = a_{PQ} Q + \xi_{PQ} G$.\\
In fact, let us observe that, in the more general case that we are considering, the $U$-dependent anomalous term \eqref{U-em-anomaly} is no more invariant under the $U(1)_{PQ}$ transformation \eqref{U(1)_PQ_new}, but it changes (under an infinitesimal transformation) as
\begin{equation}
U(1)_{PQ}:~~~~ \Delta\mathcal{L}_{U,\text{anomaly}}^{\text{(e.m.)}} \rightarrow \Delta\mathcal{L}_{U,\text{anomaly}}^{\text{(e.m.)}} - \gamma \Tr[\mathcal{Q}_{e.m.}^2 \mathcal{Q}_{PQ}] \, G .
\end{equation}
Therefore, since under the same $U(1)_{PQ}$ transformation the $N$-dependent anomalous term \eqref{N-em-anomaly} transforms as
\begin{equation}
U(1)_{PQ}:~~~~ \Delta\mathcal{L}_{N,\text{anomaly}}^{\text{(e.m.)}} \rightarrow \Delta\mathcal{L}_{N,\text{anomaly}}^{\text{(e.m.)}} - \gamma \tilde\xi_{PQ} \, G ,
\end{equation}
it comes out that, with $\tilde\xi_{PQ}$ defined in Eq. \eqref{xi-tilde_PQ}, the total electromagnetic anomalous term
$\Delta\mathcal{L}_{\text{anomaly}}^{\text{(e.m.)}} = \Delta\mathcal{L}_{U, \text{anomaly}}^{\text{(e.m.)}} + \Delta\mathcal{L}_{N, \text{anomaly}}^{\text{(e.m.)}}$ trasforms in the correct way, i.e.,
\begin{equation}
U(1)_{PQ}:~~~~ \Delta\mathcal{L}_{\text{anomaly}}^{\text{(e.m.)}} \rightarrow \Delta\mathcal{L}_{\text{anomaly}}^{\text{(e.m.)}} - \gamma \xi_{PQ} \, G .
\end{equation}
Making use of Eq. \eqref{mixing}, one immediatly sees that this term contains an axion-photon-photon interaction of the type
\begin{equation}\label{gammagammacoupling}
\Delta\mathcal{L}_{a\gamma\gamma} = -\frac{1}{4}g_{a\gamma\gamma}aF_{\mu\nu}\tilde{F}^{\mu\nu} ,
\end{equation}
with the following expression for the \emph{axion-photon-photon coupling constant}:
\begin{equation}\label{gammagamma}
g_{a\gamma\gamma}= \frac{\alpha_{\text{e.m.}}}{\pi F_{\pi}}\left(\theta_{\pi_{3}S_a}+\frac{1}{\sqrt{3}}\theta_{\pi_{8}S_a}+\frac{2\sqrt{2}}{\sqrt{3}}\theta_{SS_a}\right) + \frac{\alpha_{\text{e.m.}}}{\pi f_a} 3\tilde\xi_{PQ}\theta_{S_a S_a} ,
\end{equation}
where $\alpha_{\text{e.m.}} = \frac{e^2}{4\pi} \simeq \frac{1}{137}$ is the fine-structure constant.

The columns of the orthogonal mixing matrix $\theta_{ij}$ in Eq. \eqref{mixing} are just the eigenvectors of the squared-mass matrix \eqref{squared-mass-matrix}. In particular, the four mixing parameters appearing in Eq. \eqref{gammagamma} are just the components of the eigenvector $\Ket{a}$ corresponding to the ``physical'' axion and they can thus be determined by solving the following equation:
\begin{equation}\label{Physical_Axion}
\Bigl(\mathcal{M}^2 - m_a^2 \, \mathbf{I}\Bigr) \Ket{a} = 0 ,~~~~\text{where:}~~
\Ket{a} =
\begin{pmatrix}
\theta_{\pi_3 S_a} \\
\theta_{\pi_8 S_a} \\
\theta_{S S_a} \\
\theta_{S_a S_a} \\
\end{pmatrix}
\end{equation}
and $m_a^2$ is the axion squared-mass computed in the previous section at the leading order in $f_a^{-1}$, see Eq. \eqref{Axion_Mass}.
We aim to compute the meson-axion mixing parameters \eqref{Physical_Axion} at the first order in $f_a^{-1}$: therefore, since the axion squared-mass $m_a^2$ turns out to be of order $\mathcal{O}(f_a^{-2})$, we can basically neglect it and take the approximation $m_a^2 \simeq 0$. In other words, we thus look for the kernel of $\mathcal{M}^2$ (i.e., for the solutions of $\mathcal{M}^2 \Ket{a} = 0$). \\
Accordingly, Eq. \eqref{Physical_Axion} reads:
\begin{equation}\label{Equations_Axion_Mixing_Parameters}
\begin{cases}
2 B \tilde{m} \, \theta_{\pi_3 S_a} + \frac{1}{\sqrt{3}} B \Delta \, \theta_{\pi_8 S_a} + \sqrt{\frac{2}{3}} B \Delta \, \theta_{S S_a} - \frac{B F_\pi}{f_a} m_{p_1} \, \theta_{S_a S_a} = 0 ,\\
	\displaystyle \frac{1}{\sqrt{3}} B \Delta \, \theta_{\pi_3 S_a} + \frac{2}{3}B \left(\tilde{m} + 2m_s\right) \, \theta_{\pi_8 S_a} + \frac{2\sqrt{2}}{3}B \left(\tilde{m} - m_s\right) \, \theta_{S S_a} - \frac{B F_\pi}{\sqrt{3} f_a} m_{p_2} \, \theta_{S_a S_a} = 0 ,\\
\begin{split}
\sqrt{\frac{2}{3}} B \Delta \, \theta_{\pi_3 S_a} + \frac{2\sqrt{2}}{3}B \left(\tilde{m} - m_s\right) \, \theta_{\pi_8 S_a} & + \bigg[\frac{6A}{F_{\pi}^2} + \frac{2}{3}B \big(2\tilde{m} + m_s\big)\bigg] \, \theta_{S S_a} \\
& + \biggl(\frac{\sqrt{6} A \tilde{a}_{PQ}}{F_\pi f_a} - \sqrt{\frac{2}{3}}\frac{B F_\pi}{f_a} m_{p_3}\biggr) \, \theta_{S_a S_a} = 0 ,
\end{split}
\\
\displaystyle \theta_{\pi_3 S_a}^2 + \theta_{\pi_8 S_a}^2 + \theta_{S S_a}^2 + \theta_{S_a S_a}^2 = 1 .
\end{cases}
\end{equation}
The last equation is the normalization condition for the column vectors of the orthogonal matrix $\theta_{i j}$: it can be easily satisfied (at the first order in $f_a^{-1}$) by taking $\theta_{S_a S_a} = 1$, being the other mixing parameters $\theta_{\pi_3 S_a}$, $\theta_{\pi_8 S_a}$, and $\theta_{S S_a}$ of order $\mathcal{O}(f_a^{-1})$ (so that their squared values are of order $\mathcal{O}(f_a^{-2})$ and thus negligible).\\
The meson-axion mixing coefficients have been computed with the help of \textit{Mathematica} and their expressions are given by:
\begin{align}\label{Axion_Mixing_Parameters}
\theta_{\pi_3 S_a} & = \frac{F_{\pi}}{2 f_a} \left\{ \frac{\tilde{a}_{PQ} (m_u - m_d) m_s}{(m_u + m_d) m_s + m_u m_d \Big(1 + \frac{B F_{\pi}^2}{A} m_s\Big)} \right. \nonumber \\
& + \left. \frac{(m_u - m_d) m_s p_s + 2 m_s (m_u p_u - m_d p_d) + m_u m_d (p_u - p_d) \Big(1 + \frac{B F_{\pi}^2}{A} m_s\Big)}{(m_u + m_d) m_s + m_u m_d \Big(1 + \frac{B F_{\pi}^2}{A} m_s\Big)} \right\} ,\\
\theta_{\pi_8 S_a} & = - \frac{F_{\pi}}{2\sqrt{3} f_a} \left\{ \frac{\tilde{a}_{PQ} [(m_u + m_d) m_s - 2 m_u m_d]}{(m_u + m_d) m_s + m_u m_d \Big(1 + \frac{B F_{\pi}^2}{A} m_s\Big)} \right. \nonumber \\
& + \left. \frac{3 (m_u + m_d) m_s p_s - 3 m_u m_d (p_u + p_d) - \frac{B F_{\pi}^2}{A} m_u m_d m_s (p_u + p_d - 2 p_s)}{(m_u + m_d) m_s + m_u m_d \Big(1 + \frac{B F_{\pi}^2}{A} m_s\Big)} \right\} ,\\
\theta_{S S_a} & = - \frac{F_{\pi}}{\sqrt{6} f_a} \left\{ \frac{\tilde{a}_{PQ} [(m_u + m_d) m_s + m_u m_d] - \frac{B F_{\pi}^2}{A} m_u m_d m_s (p_u + p_d + p_s)}{(m_u + m_d) m_s + m_u m_d \Big(1 + \frac{B F_{\pi}^2}{A} m_s\Big)} \right\} ,\\
\theta_{S_a S_a} & = 1 .
\end{align}
Then, substituting the above results into Eq. \eqref{gammagamma} and using also the expressions \eqref{a-tilde_PQ} and \eqref{xi-tilde_PQ} for the \textit{reduced} color and electromagnetic anomaly oefficients $\tilde{a}_{PQ}$ and $\tilde\xi_{PQ}$, we obtain the following expression for the axion-photon-photon coupling constant:
\begin{equation}\label{axion-photon-coupling}
g_{a \gamma \gamma} = \frac{\alpha_{e.m.}}{\pi} \left(\frac{a_{PQ}}{f_a}\right) \left\{ \frac{3\xi_{PQ}}{a_{PQ}} - \frac{1}{3} \frac{(m_u + 4 m_d)m_s + m_u m_d}{(m_u + m_d) m_s + m_u m_d \Big(1 + \frac{B F_{\pi}^2}{A} m_s\Big)} \right\} .
\end{equation}
Therefore, in agreement with a well-known result obtained in the literature using a different method, we have found that the axion-photon-photon coupling constant is the sum of two contributions,
$g_{a \gamma \gamma} = g^{(0)}_{a \gamma \gamma} + g^{(1)}_{a \gamma \gamma}$,
where $g^{(0)}_{a\gamma\gamma} \equiv \frac{\alpha_{e.m.}}{\pi} \left(\frac{a_{PQ}}{f_a}\right) \frac{3\xi_{PQ}}{a_{PQ}}$ is a model-dependent contribution proportional to the electromagnetic anomaly of the $U(1)_{PQ}$ current, while
\begin{equation}\label{g^(1)}
g^{(1)}_{a \gamma \gamma} \equiv -\frac{\alpha_{e.m.}}{3\pi} \left(\frac{a_{PQ}}{f_a}\right) \frac{(m_u + 4 m_d)m_s + m_u m_d}{(m_u + m_d) m_s + m_u m_d \Big(1 + \frac{B F_{\pi}^2}{A} m_s\Big)}
\end{equation}
is (apart from the multiplicative factor $a_{PQ}/f_a$) a model-independent contribution, which depends on the light quark masses and on some low-energy constants of QCD.\footnote{In the literature (see, e.g., Refs. \cite{DMN2017,GHVV2016} and references therein) the coefficients of the \emph{color anomaly} and of the \emph{electromagnetic anomaly} are sometimes denoted as ``$N$'' and ``$E$'' respectively and they are defined through the relation:
$\partial_{\mu}J^{\mu}_{PQ} = N \frac{\alpha_s}{4\pi} G^{a}_{\mu\nu} \tilde{G}_{a}^{\mu\nu} + E \frac{\alpha_{\text{e.m.}}}{4\pi} F_{\mu\nu} \tilde{F}^{\mu\nu} = 2N Q + \frac{1}{3}E G$,
being $Q=\frac{g^{2}}{64\pi^{2}}\varepsilon^{\mu\nu\rho\sigma} G^{a}_{\mu\nu}G^{a}_{\rho\sigma} = \frac{\alpha_s}{8\pi} G^{a}_{\mu\nu} \tilde{G}_{a}^{\mu\nu}$,
where $\alpha_s = \frac{g^2}{4\pi}$,
and $G=\frac{e^2 N_C}{32\pi^2}\varepsilon^{\mu\nu\rho\sigma}F_{\mu\nu}F_{\rho\sigma} = \frac{3 \alpha_{\text{e.m.}}}{4\pi} F_{\mu\nu} \tilde{F}^{\mu\nu}$,
where $\alpha_{\text{e.m.}} = \frac{e^2}{4\pi}$ is the fine-structure constant.
Therefore, the coefficients $a_{PQ}$ and $\xi_{PQ}$, that we have defined in this paper, are related to the coefficients $N$ and $E$ by the following relations: $a_{PQ} = 2N$ and $\xi_{PQ} = \frac{1}{3} E$.
Moreover, it is customary in the literature to define the energy scale $f_a$ at which the $U(1)_{PQ}$ global symmetry is broken so as to coincide with our $f_a/a_{PQ}$.
}
\\
Eqs. \eqref{axion-photon-coupling}--\eqref{g^(1)} represent an improvement of the corresponding result found in Ref. \cite{LM2020} (using an ``axionized'' version of the WDV model in which the light quarks are neutral under the $U(1)_{PQ}$ symmetry), where only approximate expressions for the meson-axion mixing parameters and for the model-independent contribution $g^{(1)}_{a \gamma \gamma}$ (denoted there as ``$g^{QCD}_{a \gamma \gamma}$'') were derived at the first order in an expansion in powers of the ``small'' parameter $\Delta \equiv m_u - m_d$.\\
Moreover, we observe that, if one integrates out the \emph{strange} quark, by taking the formal limits $m_s \to \infty$ and $A \to \infty$ (in the real world $B m_u, \, B m_d \ll B m_s \approx \frac{A}{F_\pi^2}$, as one can see from the numerical values reported below), the expression that we have found for the model-independent contribution $g^{(1)}_{a\gamma\gamma}$ in Eq. \eqref{g^(1)} correctly reduces to the corresponding expression derived using the Chiral Effective Lagrangian ($\chi EL$) with $N_f=2$ light quark flavors at the leading order (LO) in the momentum expansion [$\mathcal{O}(p^2)$] (see Ref. \cite{GHVV2016} and references therein):
\begin{equation}\label{g^(1)_chiEL_LO}
g^{(1)}_{a \gamma \gamma}\vert_{\chi EL}^{(\text{LO})} = -\frac{\alpha_{e.m.}}{3\pi} \left(\frac{a_{PQ}}{f_a}\right) \left( \frac{m_u + 4 m_d}{m_u + m_d} \right) .
\end{equation}
We can obtain a numerical estimate for the model-independent axion-photon-photon coupling constant \eqref{g^(1)} using the following values of the known parameters:
\begin{itemize}
\item $F_\pi=(92.1 \pm 0.9)$ MeV (see Ref. \cite{PDG}, where the value of $f_{\pi}\equiv\sqrt{2}F_{\pi}$ is reported).
\item $A=(180 \pm 5 ~\text{MeV})^4$ (see Ref. \cite{VP2009} and references therein).
\item For what concerns the parameter $B$ and the quark masses $m_{u}$, $m_{d}$, $m_{s}$, we can make use of the well-known relations (see, e.g., Ref. \cite{Weinberg-book}) between $Bm_u$, $Bm_d$, $Bm_s$ and the squared masses of the pseudoscalar mesons, derived using leading-order chiral perturbation theory (and ignoring small corrections due to the mixing with the axion):
\begin{equation}\label{masse}
\begin{cases}
Bm_{u} = m_{\pi^{0}}^{2}-\frac{1}{2}(m_{K^{0}}^{2}-m_{K^{+}}^{2}+m_{\pi^{+}}^{2}) ,\\
Bm_{d} = \frac{1}{2}(m_{K^{0}}^{2}-m_{K^{+}}^{2}+m_{\pi^{+}}^{2}) ,\\
Bm_{s} = \frac{1}{2}(m_{K^{0}}^{2}+m_{K^{+}}^{2}-m_{\pi^{+}}^{2}) .
\end{cases}
\end{equation}
The masses of the pseudoscalar mesons are given by \cite{PDG}
\begin{equation}\label{masses}
\begin{cases}
m_{\pi^{+}}=139.57039(18) ~\text{MeV} ,\\
m_{\pi^{0}}=134.9768(5) ~\text{MeV} ,\\
m_{K^{+}}=493.677(15) ~\text{MeV} ,\\
m_{K^{0}}=497.611(13) ~\text{MeV} .
\end{cases}
\end{equation}
\end{itemize}
In Table 1 we report the numerical estimate for the model-independent axion-photon-photon coupling constant $g^{(1)}_{a\gamma\gamma}$ \big[divided by the factor $b \equiv \frac{F_\pi}{\sqrt{2}}\big(\frac{a_{PQ}}{f_a}\big)$\big], obtained using the expression \eqref{g^(1)} that we have derived above using the ``axionized'' WDV model: for comparison, we also report the corresponding estimates derived using the Chiral Effective Lagrangian ($\chi EL$) with $N_f=2$ light quark flavors at LO [$\mathcal{O}(p^{2})$] and NLO [$\mathcal{O}(p^{4})$] (see Ref. \cite{GHVV2016} and references therein).
The numerical value of $g^{(1)}_{a \gamma \gamma}$ that we have obtained using the expression \eqref{g^(1)} is in agreement (within the errors) with the corresponding result found in Ref. \cite{LM2020}, using an approximate expression which was derived at the first order in an expansion in powers of the ``small'' parameter $\Delta \equiv m_u - m_d$. Moreover, it is also consistent with the prediction of the Chiral Effective Lagrangian ($N_f = 2$) at NLO [$\mathcal{O}(p^4)$].
\begin{table}
\begin{center}
\begin{tabular}{| l | l |}
\hline
& $|g^{(1)}_{a\gamma\gamma}|/b$ [$\text{MeV}^{-1}$] \\ \hline
$\chi EL$ ($N_f=2$) at LO \cite{GHVV2016} & $(3.59\pm0.05) \times 10^{-5}$ \\ \hline
$\chi EL$ ($N_f=2$) at NLO \cite{GHVV2016} & $(3.42\pm0.07) \times 10^{-5}$ \\ \hline
WDV ($N_f=3$) [Eq. \eqref{g^(1)}] & $(3.33\pm0.04) \times 10^{-5}$ \\ \hline
\end{tabular}
\end{center}
\caption{Numerical values of $|g^{(1)}_{a\gamma\gamma}|/b$, with $b \equiv \frac{F_\pi}{\sqrt{2}}\big(\frac{a_{PQ}}{f_a}\big)$, obtained using the expression \eqref{g^(1)}, compared with the predictions of the Chiral Effective Lagrangian ($N_f=2$) at LO and NLO.}
\end{table}

\section{Hadronic decays with the axion}
\setcounter{equation}{0}

In Ref. \cite{LM2020} the following processes (involving the lightest hadrons, i.e., the pseudoscalar mesons, and the axion),
\begin{equation}\label{Hadronic_Decays}
\begin{cases}
\eta \rightarrow \pi^0 + \pi^0 + a ,\\
\eta \rightarrow \pi^+ + \pi^- + a ,\\
\eta' \rightarrow \pi^0 + \pi^0 + a ,\\
\eta' \rightarrow \pi^+ + \pi^- + a ,\\
\end{cases}
\end{equation}
were studied using the ``axionized'' version \eqref{WDV+axion_old} of the WDV model, where the light quarks are neutral under the $U(1)_{PQ}$ transformation.
In this section we shall extend this study using, instead, the new ``axionized'' version \eqref{WDV+axion} of the WDV model, which allows for the possibility that the light quarks may have nonzero PQ charges.
As it was also remarked in Ref. \cite{LM2020}, the couplings of the axion with hadrons in general (and with the lightest mesons in particular) had already been investigated in the past literature, in many cases using also chiral effective Lagrangian techniques (see, for instance, Refs. \cite{BPY1987,GKR1986,KW1986}), but the particular processes \eqref{Hadronic_Decays} had not been explicitly investigated in those previous works.\footnote{However, in Ref. \cite{ASW2019} (and, more recently, also in Ref. \cite{OZ2025}) similar processes, such as $a\rightarrow 3\pi$ or $a\rightarrow \eta(\eta')\pi\pi$, involving GeV-scale \emph{axionlike} particles, were investigated, using chiral effective Lagrangian techniques and also incorporating heavier resonances, which are relevant at the GeV scale.}
More recently, the processes \eqref{Hadronic_Decays} have been also studied in Ref. \cite{AG2024}, using a chiral effective Lagrangian framework very similar to the one adopted in this paper and, moreover, including also (by means of dispersion relations) the rescattering effects of the final-state pions.

In order to compute the amplitudes of the processes \eqref{Hadronic_Decays}, we must expand the ``axionized'' WDV Lagrangian \eqref{WDV+axion} up to the fourth order in powers of the pseudoscalar fields (both mesons and axion), thereby obtaining the interaction vertices corresponding to the above-reported processes.
We thus obtain the following quartic Lagrangian:
\begin{equation}\label{Quartic_Lagrangian_Full}
\begin{split}
\mathcal{L}^{(4)} & = \frac{1}{4 F_{\pi}^2} \bigg[- \frac{2}{3} f_{i j c} f_{c \alpha \beta} \left(\pi_i \partial_{\mu} \pi_j \right) \left(\pi_{\alpha} \partial^{\mu} \pi_{\beta} \right)\bigg] + \frac{B}{24 F_{\pi}^2} \; \text{Tr} \Big[M \; \Phi^4\Big] \\
& - \frac{B}{6 F_{\pi} f_a} \; \text{Tr} \bigg[M \; \Bigl\{\Phi^3, \mathcal{Q}_{PQ} \Bigr\}\bigg] S_a + \frac{B}{4 f_a^2} \; \text{Tr} \bigg[M \; \Bigl\{\Phi^2, \mathcal{Q}_{PQ}^2 \Bigr\}\bigg] S_a^2 \\
& - \frac{B F_{\pi}}{6 f_a^3} \; \text{Tr} \bigg[M \; \Bigl\{\Phi, \mathcal{Q}_{PQ}^3 \Bigr\}\bigg] S_a^3 + \frac{B F_{\pi}^2}{24 f_a^4} \; \text{Tr} \Big[M \; \mathcal{Q}_{PQ}^4 \Big] S_a^4 ,
\end{split}
\end{equation}
where $f_{abc}$ are the $SU(3)$ structure constants (defined as $[\lambda_{a},\lambda_b]=2if_{abc}\lambda_c$), $\{A,B\}$ is the anticommutator $AB + BA$, and $\Phi(x)$ is the Hermitian matrix field appearing at the exponent of the unitary matrix field $U = e^{\frac{i}{F_\pi} \Phi}$ [see Eq. \eqref{U-field}].\\
It is easy to see that the first term in the right-hand side of Eq. \eqref{Quartic_Lagrangian_Full} does not reproduce any of the processes \eqref{Hadronic_Decays}.\\
Analogously to what we have done in the previous sections, we shall compute the decay amplitudes only at the leading order in $f_a^{-1}$ and, moreover, we are interested in processes involving only one axion, so that we shall neglect the last three terms in Eq. \eqref{Quartic_Lagrangian_Full}.\\
In the following, we shall also consider (as it was done in Ref. \cite{LM2020}) the approximation $\Delta = 0$, thus neglecting the small mass difference between the \textit{up} and \textit{down} quarks ($m_u = m_d = \tilde{m}$).
Accordingly, we diagonalize the squared-mass matrix \eqref{squared-mass-matrix} at the first order in the energy scale $f_a^{-1}$ within the approximation $\Delta = 0$ and [using also the fact that, as it has been already remarked in Sect. 3, the eigenvalues of the squared-mass matrix \eqref{squared-mass-matrix} get corrections of order $\mathcal{O}(f_a^{-2})$ with respect to their ``unperturbed'' values at $f_a\to\infty$, i.e., when the axion is decoupled] the following mixing parameters are obtained:
\begin{equation}\label{mixing-parameters}
\begin{cases}
\theta_{\pi_3 \pi_3} = 1,~~ \theta_{\pi_3 \pi_8} = \theta_{\pi_3 S} = 0,~~ \theta_{\pi_3 S_a} = \mathcal{O}(f_a^{-1}) ,\\
\theta_{\pi_8 \pi_3} = 0,~~ \theta_{\pi_8 \pi_8} = \cos\varphi,~~ \theta_{\pi_8 S} = -\sin\varphi,~~ \theta_{\pi_8 S_a} = \mathcal{O}(f_a^{-1}) ,\\
\theta_{S \pi_3} = 0,~~ \theta_{S \pi_8} = \sin\varphi,~~ \theta_{S S} = \cos\varphi,~~ \theta_{S S_a} = \mathcal{O}(f_a^{-1}) ,\\
\theta_{S_a \pi_3} = \mathcal{O}(f_a^{-1}),~~ \theta_{S_a \pi_8} = \mathcal{O}(f_a^{-1}),~~ \theta_{S_a S} = \mathcal{O}(f_a^{-1}),~~ \theta_{S_a S_a} = 1 ,
\end{cases}
\end{equation}
where $\varphi$ is the mixing angle between $\pi_8$ and $S$, given by \cite{Veneziano1979}:
\begin{equation}\label{mixing-angle}
\tan\varphi = \sqrt{2} - \frac{3}{2\sqrt{2}} \left[ \frac{m_\eta^2 - 2B\tilde{m}}{B(m_s-\tilde{m})} \right] .
\end{equation}
The explicit expressions for the meson-axion mixing coefficients $\theta_{\pi_3 S_a}$, $\theta_{\pi_8 S_a}$ and $\theta_{S S_a}$ have been derived in the previous section (when studying the electromagnetic decay of the axion). We report here the results for seek of clarity, by imposing the approximation $\Delta = 0$:
\begin{align}
\theta_{\pi_3 S_a} & = \frac{F_{\pi}}{2 f_a} (p_u - p_d) , \\
\theta_{\pi_8 S_a} & = - \frac{F_{\pi}}{2\sqrt{3} f_a} \left\{ \frac{2 \tilde{a}_{PQ} (m_s - \tilde{m}) + 6 m_s p_s - 3 \tilde{m} (p_u + p_d) - \frac{B F_{\pi}^2}{A} \tilde{m} m_s (p_u + p_d - 2 p_s)}{2 m_s + \tilde{m} \Big(1+ \frac{B F_{\pi}^2}{A} m_s\Big)} \right\} , \\
\theta_{S S_a} & = - \frac{F_{\pi}}{\sqrt{6} f_a} \left\{ \frac{\tilde{a}_{PQ} \big(\tilde{m} + 2 m_s\big) - \frac{B F_{\pi}^2}{A} \tilde{m} m_s (p_u + p_d + p_s)}{2 m_s + \tilde{m} \Big(1 + \frac{B F_{\pi}^2}{A} m_s\Big)} \right\} .
\end{align}
At the end, the relevant interaction Lagrangian describing the aforementioned hadronic decays \eqref{Hadronic_Decays} turns out to be:
\begin{equation}\label{Quartic_Interaction_Lagrangian}
\Delta \mathcal{L}^{(4)} = \frac{B \tilde{m}}{3 F_{\pi}^2} \biggl(\frac{1}{2} \pi_3^2 + \pi^+ \pi^-\biggr)
\Big[\pi_8^2 + 2 S^2 + 2 \sqrt{2} \pi_8 S + \frac{2 \sqrt{6} F_{\pi}}{f_a} \left(p_u + p_d \right) \big(\sqrt{2} \pi_8 S_a + S S_a\big)\Big] ,
\end{equation}
where $\pi^{\pm}=\frac{\pi_1\mp i\pi_2}{\sqrt{2}}$ are the charged pion fields.\\
After expressing the fields $\pi_3$, $\pi_8$, $S$, and $S_a$ in terms of the \emph{physical} fields $\pi^0$, $\eta$, $\eta'$, and $a$ by means of Eq. \eqref{mixing} and making use of the above-reported expressions of the mixing parameters, the following quartic interaction terms are derived:
\begin{equation}\label{Interaction_Vertices_Hadronic_Decays}
\begin{cases}
\Delta\mathcal{L}_{\eta\pi^{0}\pi^{0}a}=\frac{1}{2}g_{\eta\pi^{0}\pi^{0}a}\eta(\pi^{0})^2 a ,\\
\Delta\mathcal{L}_{\eta\pi^{+}\pi^{-}a}=g_{\eta\pi^{+}\pi^{-}a}\eta \pi^{+}\pi^{-}a ,\\
\Delta\mathcal{L}_{\eta'\pi^{0}\pi^{0}a}=\frac{1}{2}g_{\eta'\pi^{0}\pi^{0}a}\eta'(\pi^{0})^2 a ,\\
\Delta\mathcal{L}_{\eta'\pi^{+}\pi^{-}a}=g_{\eta'\pi^{+}\pi^{-}a}\eta'\pi^{+}\pi^{-} a ,
\end{cases}
\end{equation}
where
\begin{equation}\label{Axion_Amplitude_1}
\begin{split}
g_{\eta \pi^0 \pi^0 a} & = g_{\eta \pi^+ \pi^- a} = - \frac{2 B \tilde{m}}{\sqrt{3} F_{\pi} f_a} \big( \cos\varphi + \sqrt{2} \sin\varphi \big) \\
& \times \Bigg\{ \tilde{a}_{PQ} \Bigg[ \frac{m_s}{2m_s + \tilde{m} \big(1 + \frac{B F_{\pi}^2}{A} m_s\big)} \Bigg] + \frac{1}{4} \Bigg[ \frac{2 m_s p_s - \tilde{m} \big(1 + \frac{B F_{\pi}^2}{A} m_s\big) \big(p_u + p_d\big)}{2m_s + \tilde{m} \big(1 + \frac{B F_{\pi}^2}{A} m_s\big)} \Bigg] \\
& - \sqrt{2} \bigg( \frac{\sqrt{2} + \tan\varphi}{1 + \sqrt{2}\tan\varphi} \bigg) \big(p_u + p_d\big) \Bigg\} \\
& = - \frac{B \tilde{m}}{\sqrt{3} F_{\pi}} \left(\frac{a_{PQ}}{f_a}\right) \big( \cos\varphi + \sqrt{2} \sin\varphi \big) \,
\Bigg\{ \frac{1}{1 + \frac{\tilde{m}}{2 m_s} \left( 1 + \frac{B F_{\pi}^2}{A} m_s \right)} \\
& \times \left[ 1 - \bigg(1 + \frac{\tilde{m}}{4 m_s} \Big( 1 + \frac{B F_{\pi}^2}{A} m_s \Big)\bigg) \bigg(\frac{p_u + p_d}{a_{PQ}}\bigg) - \frac{1}{2}\bigg(\frac{p_s}{a_{PQ}}\bigg) \right] \\
& - 2\sqrt{2} \bigg( \frac{\sqrt{2} + \tan\varphi}{1 + \sqrt{2}\tan\varphi}\bigg) \bigg(\frac{p_u + p_d}{a_{PQ}}\bigg) \Bigg\}
\end{split}
\end{equation}
and
\begin{equation}\label{Axion_Amplitude_2}
\begin{split}
g_{\eta' \pi^0 \pi^0 a} & = g_{\eta' \pi^+ \pi^- a} = - \frac{2 B \tilde{m}}{\sqrt{3} F_{\pi} f_a} \big( \sqrt{2} \cos\varphi - \sin\varphi \big) \\
& \times \Bigg\{ \tilde{a}_{PQ} \Bigg[ \frac{m_s}{2m_s + \tilde{m} \big(1 + \frac{B F_{\pi}^2}{A} m_s\big)} \Bigg] + \frac{1}{4} \Bigg[ \frac{2 m_s p_s - \tilde{m} \big(1 + \frac{B F_{\pi}^2}{A} m_s\big) \big(p_u + p_d\big)}{2m_s + \tilde{m} \big(1 + \frac{B F_{\pi}^2}{A} m_s\big)} \Bigg] \\
& - \sqrt{2} \bigg( \frac{1 - \sqrt{2} \tan\varphi}{\sqrt{2} - \tan\varphi}\bigg) \big(p_u + p_d\big) \Bigg\} \\
& = - \frac{B \tilde{m}}{\sqrt{3} F_{\pi}} \left(\frac{a_{PQ}}{f_a}\right) \big( \sqrt{2}\cos\varphi - \sin\varphi \big) \,
\Bigg\{ \frac{1}{1 + \frac{\tilde{m}}{2 m_s} \left( 1 + \frac{B F_{\pi}^2}{A} m_s \right)} \\
& \times \left[ 1 - \bigg(1 + \frac{\tilde{m}}{4 m_s} \Big( 1 + \frac{B F_{\pi}^2}{A} m_s \Big)\bigg) \bigg(\frac{p_u + p_d}{a_{PQ}}\bigg) - \frac{1}{2}\bigg(\frac{p_s}{a_{PQ}}\bigg) \right] \\
& - 2\sqrt{2} \bigg( \frac{1 - \sqrt{2} \tan\varphi}{\sqrt{2} - \tan\varphi} \bigg) \bigg(\frac{p_u + p_d}{a_{PQ}}\bigg) \Bigg\} .
\end{split}
\end{equation}
The decay amplitude for each of the four processes \eqref{Hadronic_Decays} is nothing but the coupling constant $g$ in the corresponding interaction term in Eq. \eqref{Interaction_Vertices_Hadronic_Decays}.
Accordingly, the decay widths for the processes \eqref{Hadronic_Decays} are given by
\begin{equation}\label{Decay_Widths}
\begin{cases}
\Gamma(\eta\rightarrow\pi^{0}\pi^{0}a)=\frac{1}{2m_{\eta} \cdot 2!}|g_{\eta\pi^{0}\pi^{0}a}|^{2}\Phi^{(3)}(m_{\eta}|m_{\pi^{0}},m_{\pi^{0}},m_a) ,\\
\Gamma(\eta\rightarrow\pi^{+}\pi^{-}a)=\frac{1}{2m_{\eta}}|g_{\eta\pi^{+}\pi^{-}a}|^{2}\Phi^{(3)}(m_{\eta}|m_{\pi^{+}},m_{\pi^{-}},m_a) ,\\
\Gamma(\eta'\rightarrow\pi^{0}\pi^{0}a)=\frac{1}{2m_{\eta'} \cdot 2!}|g_{\eta'\pi^{0}\pi^{0}a}|^{2}\Phi^{(3)}(m_{\eta'}|m_{\pi^{0}},m_{\pi^{0}},m_a) ,\\
\Gamma(\eta'\rightarrow\pi^{+}\pi^{-}a)=\frac{1}{2m_{\eta'}}|g_{\eta'\pi^{+}\pi^{-}a}|^{2}\Phi^{(3)}(m_{\eta'}|m_{\pi^{+}},m_{\pi^{-}},m_a) ,
\end{cases}
\end{equation}
where $\Phi^{(3)}(M|m_1,m_2,m_3)$ is the phase space (with the usual ``relativistic'' normalization) for three particles of masses $m_1$, $m_2$, $m_3$ with total energy $M$ in the center-of-mass system. The exact expression is rather complicated (see Eq. (3.18) in Ref. \cite{Meggiolaro2011}, and also Ref. \cite{DD2004} and references therein), but it is surely a good approximation to take $m_a\simeq0$, considering the experimental upper bound on the axion mass $m_a\lesssim 10^{-2}$ eV \cite{bounds_a1,bounds_a2,bounds_b,IR2018}. The expression for the phase space for two particles of mass \textit{m} and one massless particle turns out to be
\begin{equation}
\begin{split}
&\Phi^{(3)}(M|m,m,0) = \\
&\frac{M^{2}}{256\pi^{3}}\left\{\left(1+\frac{2m^{2}}{M^{2}}\right)\sqrt{1-\frac{4m^{2}}{M^{2}}}-\frac{4m^{2}}{M^{2}}\left(1-\frac{m^{2}}{M^{2}}\right)\ln\left[\frac{M^{2}}{2m^{2}}\left(1+\sqrt{1-\frac{4m^{2}}{M^{2}}}\right)-1\right]\right\} .
\end{split}
\end{equation}
We notice that, differently from what we have found in the previous sections concerning the axion mass [see Eq. \eqref{Axion_Mass_Chi}] and the axion-photon-photon coupling constant [see Eqs. \eqref{axion-photon-coupling}--\eqref{g^(1)}], the results for the coupling constants \eqref{Axion_Amplitude_1} and \eqref{Axion_Amplitude_2}, corresponding to the decay amplitudes of the processes \eqref{Hadronic_Decays}, strongly depend on the details of the model, i.e., on the values of the PQ charges $p_u$, $p_d$, and $p_s$.\\
In order to put in evidence this strong model dependence, we shall report the expressions for the coupling constants (i.e., the amplitudes) \eqref{Axion_Amplitude_1} and \eqref{Axion_Amplitude_2} and we shall evaluate numerically the corresponding decay widths in the two following \emph{extreme} cases: (i) the case in which $\mathcal{Q}_{PQ} = 0$, so that the color anomaly parameter $a_{PQ}$ is all due to the \emph{heavy} degrees of freedom (which are integrated out in the effective Lagrangian), as it happens, for example, in the KSVZ model (this, as we have already said, is indeed the case that was considered in Ref. \cite{LM2020}); and (ii) the case in which the color anomaly is instead due only to the \emph{up} quark, i.e., $a_{PQ} = p_u$ and $p_d = p_s = 0$ (as it happens, for example, in the model considered in Ref. \cite{KW1986}).

\subsection{Case in which $\mathcal{Q}_{PQ} = 0$}

\noindent
Putting $p_u = p_d = p_s = 0$ in Eqs. \eqref{Axion_Amplitude_1}--\eqref{Axion_Amplitude_2}, one finds that
\begin{equation}\label{Axion_Amplitude_1_LM2020}
g_{\eta \pi^0 \pi^0 a} = g_{\eta \pi^+ \pi^- a} = -b\frac{\sqrt{2} B \tilde{m}}{\sqrt{3} F_{\pi}^2}
\Bigg\{
\frac{\cos\varphi + \sqrt{2} \sin\varphi}{1 + \frac{\tilde{m}}{2 m_s} \left( 1 + \frac{B F_{\pi}^2}{A} m_s \right)}
\Bigg\}
\end{equation}
and 
\begin{equation}\label{Axion_Amplitude_2_LM2020}
g_{\eta' \pi^0 \pi^0 a} = g_{\eta' \pi^+ \pi^- a} = -b\frac{\sqrt{2} B \tilde{m}}{\sqrt{3} F_{\pi}^2}
\Bigg\{
\frac{\sqrt{2} \cos\varphi - \sin\varphi}{1 + \frac{\tilde{m}}{2 m_s} \left( 1 + \frac{B F_{\pi}^2}{A} m_s \right)}
\Bigg\} ,
\end{equation}
where the parameter $b$ has been defined in Table 1.\\
Using the numerical values of the parameters reported at the end of the previous section on the electromagnetic decay of the axion (together with the values of the masses $m_{\eta} = 547.862(17)$ MeV and $m_{\eta'} = 957.78(6)$ MeV), one obtains the following results for the decay widths \eqref{Decay_Widths}:
\begin{equation}\label{Decay_Widths_LM2020}
\begin{cases}
\Gamma(\eta\rightarrow\pi^{0}\pi^{0}a)=b^{2}(5.62\pm0.04)\times10^{-3} ~\text{MeV} ,\\
\Gamma(\eta\rightarrow\pi^{+}\pi^{-}a)=b^{2}(10.52\pm0.07)\times10^{-3} ~\text{MeV} ,\\
\Gamma(\eta'\rightarrow\pi^{0}\pi^{0}a)=b^{2}(2.49\pm0.02)\times10^{-2} ~\text{MeV} ,\\
\Gamma(\eta'\rightarrow\pi^{+}\pi^{-}a)=b^{2}(5.01\pm0.03)\times10^{-2} ~\text{MeV} .
\end{cases}
\end{equation}
The results \eqref{Axion_Amplitude_1_LM2020}, \eqref{Axion_Amplitude_2_LM2020}, and \eqref{Decay_Widths_LM2020} coincide with the ones found in Ref. \cite{LM2020}, where the three light quarks were taken to be neutral under the $U(1)_{PQ}$ symmetry (as it happens, for example, in the KSVZ model).

\subsection{Case in which $a_{PQ} = p_u$ and $p_d = p_s = 0$}

\noindent
Substituting the above values of the PQ charges, one finds that
\begin{equation}\label{Axion_Amplitude_1_(a_PQ=p_u)}
\begin{split}
g_{\eta \pi^0 \pi^0 a} & = g_{\eta \pi^+ \pi^- a} =
b\frac{\sqrt{2} B \tilde{m}}{\sqrt{3} F_{\pi}^2} \big( \cos\varphi + \sqrt{2} \sin\varphi \big) \\
& \times \Bigg\{ \frac{\frac{\tilde{m}}{4 m_s} \left( 1 + \frac{B F_{\pi}^2}{A} m_s \right)}{1 + \frac{\tilde{m}}{2 m_s} \left( 1 + \frac{B F_{\pi}^2}{A} m_s \right)}
+ 2\sqrt{2} \bigg( \frac{\sqrt{2} + \tan\varphi}{1 + \sqrt{2}\tan\varphi}\bigg) \Bigg\}
\end{split}
\end{equation}
and
\begin{equation}\label{Axion_Amplitude_2_(a_PQ=p_u)}
\begin{split}
g_{\eta' \pi^0 \pi^0 a} & = g_{\eta' \pi^+ \pi^- a} =
b\frac{\sqrt{2} B \tilde{m}}{\sqrt{3} F_{\pi}^2} \big( \sqrt{2}\cos\varphi - \sin\varphi \big) \\
& \times \Bigg\{ \frac{\frac{\tilde{m}}{4 m_s} \left( 1 + \frac{B F_{\pi}^2}{A} m_s \right)}{1 + \frac{\tilde{m}}{2 m_s} \left( 1 + \frac{B F_{\pi}^2}{A} m_s \right)}
+ 2\sqrt{2} \bigg( \frac{1 - \sqrt{2}\tan\varphi}{\sqrt{2} - \tan\varphi} \bigg) \Bigg\} ,
\end{split}
\end{equation}
where, once again, $b$ is defined in Table 1 and, in this case, $a_{PQ} = p_u$.\\
The following numerical values of the decay widths \eqref{Decay_Widths} are obtained in this case:
\begin{equation}\label{Decay_Widths_(a_PQ=p_u)}
\begin{cases}
\Gamma (\eta \rightarrow \pi^0 \pi^0 a) = b^2 \big(8.94 \pm 0.05\big) \times 10^{-2} \; \text{MeV},\\
\Gamma (\eta \rightarrow \pi^+ \pi^- a) = b^2 \big(16.7 \pm 0.1\big) \times 10^{-2} \; \text{MeV}, \\
\Gamma (\eta' \rightarrow \pi^0 \pi^0 a) = b^2 \big(9.84 \pm 0.06\big) \times 10^{-2} \; \text{MeV},\\
\Gamma (\eta' \rightarrow \pi^+ \pi^- a) = b^2 \big(19.2 \pm 0.1\big) \times 10^{-2} \; \text{MeV} .
\end{cases}
\end{equation}
They turn out to be around one order of magnitude larger than the results \eqref{Decay_Widths_LM2020} found in the previous case (with $\mathcal{Q}_{PQ} = 0$).

\newpage

\section{Conclusions: summary of the results and prospects}
\setcounter{equation}{0}

In Sect. 2 of this paper we have extended the ``axionized'' version of the WDV chiral effective Lagragian model with $N_f=3$ light quark flavors studied in Refs. \cite{LM2020,DS2014,DRVY2017} to the more general case in which the light quarks (and, therefore, the mesons) may be charged under the $U(1)$ Peccei-Quinn symmetry.
In such a way, one is able to reproduce in the low-energy effective theory not only the class of high-energy axion models (corresponding essentially to the KSVZ model and its variants), where the Standard Model quarks are all \emph{neutral} under the $U(1)_{PQ}$ symmetry transformation and the $U(1)_{PQ}$ anomaly comes from heavy (``exotic'') quarks which have nonzero PQ charges (and which are integrated out in the low-energy effective theory), but also another class of axion models (like the DFSZ model and its variants, as well as the PQWW model and its variants), where also the Standard Model quarks are \emph{charged} under the $U(1)_{PQ}$ symmetry and each of them transforms with a $U(1)$ phase that depends on its PQ charge.

Using this generalized effective model, in this paper we have extended the results obtained in Ref. \cite{LM2020} and investigated the most interesting decay processes involving axions, photons and the lightest pseudoscalar mesons.

To start with, in Sect. 3 we have computed the axion mass at the leading order in $f_a^{-1}$ [see Eq. \eqref{Axion_Mass}], thus confirming the well-known relationship between the squared mass of the axion and the QCD topological susceptibility \cite{SVZ} [see Eq. \eqref{Axion_Mass_Chi}]:
$m^2_a = (\frac{a_{PQ}}{f_a})^2 \chi_{QCD}$.
Therefore, the squared mass of the axion turns out to be model independent, apart from the multiplicative factor $(a_{PQ}/f_a)^2$, which contains the \emph{total} color anomaly coefficient $a_{PQ}$.

In Sect. 4, we have studied the axion-photon electromagnetic interaction, introducing a term in the effective model that correctly reproduces the electromagnetic anomaly of the $U(1)$ and $SU(3)$ axial currents and of the $U(1)_{PQ}$ current: we have thus determined (after having computed also the meson-axion mixing parameters at the leading order in $f_a^{-1}$) the axion-photon-photon coupling constant $g_{a \gamma \gamma}$.
In agreement with a well-known result obtained in the literature using a different method, we have found [see Eq. \eqref{axion-photon-coupling}] that the axion-photon-photon coupling constant is the sum of two contributions,
$g_{a \gamma \gamma} = g^{(0)}_{a \gamma \gamma} + g^{(1)}_{a \gamma \gamma}$,
where $g^{(0)}_{a\gamma\gamma} \equiv \frac{\alpha_{e.m.}}{\pi} \left(\frac{a_{PQ}}{f_a}\right) \frac{3\xi_{PQ}}{a_{PQ}}$ is a model-dependent contribution proportional to the electromagnetic anomaly of the $U(1)_{PQ}$ current, while
$g^{(1)}_{a \gamma \gamma}$ is (apart from the multiplicative factor $a_{PQ}/f_a$) a model-independent contribution, given in Eq. \eqref{g^(1)}, which depends on the light quark masses and on some low-energy constants of QCD.\\
Eqs. \eqref{axion-photon-coupling}--\eqref{g^(1)} represent an improvement of the corresponding result found in Ref. \cite{LM2020} (using an ``axionized'' version of the WDV model in which the light quarks are neutral under the $U(1)_{PQ}$ symmetry), where only approximate expressions for the meson-axion mixing parameters and for the model-independent contribution $g^{(1)}_{a \gamma \gamma}$ (denoted there as ``$g^{QCD}_{a \gamma \gamma}$'') were derived at the first order in an expansion in powers of the ``small'' parameter $\Delta \equiv m_u - m_d$.
In Table 1 we have reported the numerical estimate for the model-independent axion-photon-photon coupling constant $g^{(1)}_{a\gamma\gamma}$ \big[divided by the factor $b \equiv \frac{F_\pi}{\sqrt{2}}\big(\frac{a_{PQ}}{f_a}\big)$\big], obtained using the expression \eqref{g^(1)} that we have derived above using the ``axionized'' WDV model: for comparison, we have also reported the corresponding estimates derived using the Chiral Effective Lagrangian ($\chi EL$) with $N_f=2$ light quark flavors at LO [$\mathcal{O}(p^{2})$] and NLO [$\mathcal{O}(p^{4})$] (see Ref. \cite{GHVV2016} and references therein).
The numerical value of $g^{(1)}_{a \gamma \gamma}$ that we have obtained using the expression \eqref{g^(1)} is consistent (within the errors) with the corresponding prediction of the Chiral Effective Lagrangian ($N_f = 2$) at NLO [$\mathcal{O}(p^4)$].

Finally, in Sect. 5 we have studied some hadronic decays involving the axion and the lightest pseudoscalar mesons, $\eta/\eta' \rightarrow \pi \pi a$, so generalizing the corresponding results found in Ref. \cite{LM2020}, where the above-mentioned processes were studied using the ``axionized'' version \eqref{WDV+axion_old} of the WDV model, in which the light quarks are neutral under the $U(1)_{PQ}$ transformation.
(As we have already recalled at the beginning of Sec. 5, these decays have been also recently studied in Ref. \cite{AG2024}, using a chiral effective Lagrangian framework very similar to the one adopted in this paper and, moreover, including also the rescattering effects of the final-state pions.)\\
Differently from what we have found for the axion mass [see Eq. \eqref{Axion_Mass_Chi}] and the axion-photon-photon coupling constant [see Eqs. \eqref{axion-photon-coupling}--\eqref{g^(1)}], the results for the coupling constants \eqref{Axion_Amplitude_1} and \eqref{Axion_Amplitude_2}, corresponding to the decay amplitudes of the above-mentioned processes \eqref{Hadronic_Decays}, come out to be strongly dependent on the details of the model, i.e., on the values of the PQ charges $p_u$, $p_d$, and $p_s$.
In order to put in evidence this strong model dependence, we have reported the expressions for the coupling constants (i.e., the amplitudes) \eqref{Axion_Amplitude_1} and \eqref{Axion_Amplitude_2} and we have evaluated numerically the corresponding decay widths in the two following \emph{extreme} cases: (i) the case in which $\mathcal{Q}_{PQ} = 0$, so that the color anomaly parameter $a_{PQ}$ is all due to the \emph{heavy} degrees of freedom (which are integrated out in the effective Lagrangian), as it happens, for example, in the KSVZ model (this, as we have already said, is indeed the case that was considered in Ref. \cite{LM2020}); and (ii) the case in which the color anomaly is instead due only to the \emph{up} quark, i.e., $a_{PQ} = p_u$ and $p_d = p_s = 0$ (as it happens, for example, in the model considered in Ref. \cite{KW1986}).
The results \eqref{Decay_Widths_(a_PQ=p_u)} for the decay widths in this last case come out to be around one order of magnitude larger than the results \eqref{Decay_Widths_LM2020} found in the first case (with $\mathcal{Q}_{PQ} = 0$).

We conclude by recalling that, from the astrophysical and cosmological bounds on the scale $f_a$ \cite{bounds_a1,bounds_a2,bounds_b} (or better on $f_a/a_{PQ}$, but $a_{PQ} \sim \mathcal{O}(1)$ for the most ``realistic'' axion models \cite{DMN2017}), i.e., $10^{9}~\textrm{GeV} \lesssim f_a \lesssim 10^{17}~\textrm{GeV}$, the following bounds for the adimensional quantity $b$, which is defined in Table 1, are derived: $10^{-18} \lesssim b \lesssim 10^{-10}$.
Therefore, the numerical results for the decay widths \eqref{Decay_Widths_LM2020} and \eqref{Decay_Widths_(a_PQ=p_u)} turn out to be very small (much smaller, for example, than the current experimental bounds on the CP-violating decays $\eta/\eta' \rightarrow \pi\pi$ \cite{PDG}:
$\Gamma^{exp}(\eta\rightarrow\pi^{0}\pi^{0})<4.6\times10^{-7}$ MeV,
$\Gamma^{exp}(\eta\rightarrow\pi^{+}\pi^{-})<5.8\times10^{-9}$ MeV,
$\Gamma^{exp}(\eta'\rightarrow\pi^{0}\pi^{0})<7.5\times10^{-5}$ MeV,
$\Gamma^{exp}(\eta'\rightarrow\pi^{+}\pi^{-})<3.4\times10^{-6}$ MeV).
This fact, of course, makes any experimental search for these processes an extremely difficult challenge.\\
Nevertheless, even if the axion-photon-photon electromagnetic coupling certainly remains the most promising source for some experimental signature of the axion, we believe that it would be worthwhile to pursue also the study of these decay processes $\eta/\eta'\rightarrow\pi\pi a$, along the line of research undertaken in Refs. \cite{LM2020,AG2024} and in this paper.

\newpage

\renewcommand{\Large}{\large}

\end{document}